\documentclass[article,onecolumn,notitlepage,floatix,amssymb,showpacs,amsmath, superscriptaddress,longbibliography]{revtex4-1}
\usepackage{graphicx}
\usepackage{epsfig}
\usepackage{amsmath,amssymb,amsthm}
\usepackage[utf8]{inputenc}
\usepackage[dvipsnames]{xcolor}
\usepackage[colorlinks, urlcolor=Black, linkcolor=Red,citecolor=CornflowerBlue]{hyperref}

\usepackage{soul}

\newcommand{\beqa}{\begin{eqnarray}}
\newcommand{\eeqa}{\end{eqnarray}}

\newcommand{\crea}[1]{{\hat #1}^\dagger}
\newcommand{\ann}[1]{\hat #1}
\newcommand{\bra}[1]{\langle #1\rvert}
\newcommand{\ket}[1]{\lvert#1\rangle}

\newcommand{\Curly}[1]{\left\{ #1\right\}}

\renewcommand{\boxed}[2]{\textcolor{#1}{%
\tikz[baseline={([yshift=-1ex]current bounding box.center)}] \node [rectangle, minimum width=1ex,rounded corners,draw] {\normalcolor\m@th$\displaystyle#2$};}}

\newcounter{appsection}
\newcounter{appsubsection}[appsection]

\usepackage{bookmark}

\begin{document}

\title{Critical parametric quantum sensing}
\author{R. Di Candia}
\email{rob.dicandia@gmail.com}
\thanks{These two authors contributed equally}
\affiliation{Department of Communications and Networking, Aalto University, Espoo, 02150 Finland}
\author{F. Minganti}
\email{fabrizio.minganti@gmail.com}
\thanks{These two authors contributed equally}
\affiliation{Institute of Physics, Ecole Polytechnique Fédérale de Lausanne (EPFL), CH-1015 Lausanne, Switzerland}
\author{K. V. Petrovnin}
\affiliation{QTF Centre of Excellence, Department of Applied Physics,
Aalto University School of Science, FI-00076 AALTO, Finland}
\author{G. S. Paraoanu}
\affiliation{QTF Centre of Excellence, Department of Applied Physics,
Aalto University School of Science, FI-00076 AALTO, Finland}
\author{S. Felicetti}
\email{felicetti.simone@gmail.com}
\affiliation{Istituto di Fotonica e Nanotecnologie, Consiglio Nazionale delle Ricerche (IFN-CNR), 00156  Roma, Italy}

\begin{abstract}
Critical quantum systems are a promising resource for quantum metrology applications, due to the diverging susceptibility developed in proximity of  phase transitions. 
Here, we assess the metrological power of parametric Kerr resonators undergoing driven-dissipative phase transitions. We fully characterize the quantum Fisher information for frequency estimation, and the Helstrom bound for frequency discrimination. By going beyond the asymptotic regime, we show that the Heisenberg precision can be achieved with experimentally reachable parameters. 
We design protocols that exploit the critical behavior of nonlinear resonators to enhance the precision of quantum magnetometers and the fidelity of superconducting qubit readout.
\end{abstract}

\maketitle

\section*{Introduction}
Criticality is a compelling resource, commonly used in classical sensing devices such as transition-edge detectors and bolometers~\cite{bolometer_review}. 
However, these devices do not follow optimal sensing strategies from the quantum mechanical point of view.
A promising approach to quantum sensing exploits quantum fluctuations in the proximity of the criticality to improve the measurement precision.
Despite a critical slowing down at the phase transition, theoretical analyses of many-body systems~\cite{tsang_quantum_2013,macieszczak_dynamical_2016,zanardi_quantum_2008,bina_dicke_2016,Ivanov_2020steady,fernandez-lorenzo_quantum_2017,rams_at_2018,heugel2020_quantum,ivanov_adiabatic_2013,invernizzi2008Optimal,Mirkhalaf2020,Wald2020,Salado2021,Niezgoda2021,mishra2021integrable,Fully_connected} show that critical quantum sensors can achieve the optimal scaling of precision~\cite{demkowicz-dobrzanski_chapter_2015}, both in the number of probes and in the measurement time~\cite{rams_at_2018,Fully_connected}. Furthermore, it has been shown~\cite{garbe2020} that finite-component phase transitions~\cite{bakemeier2012quantum,Ashhab2013, hwang_quantum_2015,Puebla2017,Zhu2020}---where the thermodynamic limit is replaced by a scaling of the system parameters~\cite{Casteels2017,Bartolo2016Exact,Minganti2018,peng_unified2019,felicetti2020universal}---can also be applied in sensing protocols. Surprisingly, quantum criticalities are versatile sensing resources that do not require the complexity of many-body system, as demonstrated by efficient dynamical protocols~\cite{Chu2021}, the inclusion of quantum-control methods~\cite{gietka2021} or ancillary probes~\cite{hu2021}, the design of multiparameter estimation protocols~\cite{Ivanov2020} and of a critical quantum-thermometer~\cite{xie2021}, and by first experimental implementations~\cite{liu2021}.

Finite-component critical sensors have hitherto been designed for light-matter interacting models where the atomic levels introduce a nonlinearity~\cite{garbe2021phase}. 
Despite their high experimental relevance in quantum optics and information~\cite{Menzel_pathent, Zhong_JPA,Marandi2014,Leghtas2015,Fedortchenko2017,Bruschi2017,Markov2018,Lescanne2020,  DiCandia_QI,DiCandia_covert, Fedorov_QT}, driven resonators with nonlinear photon-photon interactions have so far been overlooked for applications in critical quantum metrology.
These systems display a broad and exotic variety of critical phenomena, and their nontrivial dynamics and steady states depend on both the system and bath parameters~\cite{Bartolo2016Exact,Rota2019,Soriente21}.

Here, we introduce the critical parametric quantum sensor, a measurement apparatus based on the second-order driven-dissipative phase transition of a parametric nonlinear (Kerr) resonator. 
We apply tools of quantum parameter estimation, quantum hypothesis testing, and non-linear quantum optics to characterize the potential of this instrument for finite-component critical sensing. Our treatment uses the analytical solutions of the driven-dissipative Kerr resonator model~\cite{MingantiSciRep16,Bartolo2016,RobertsPRX20}, together with exact numerical calculations to: (i) Evaluate the quantum Fisher information (QFI) for frequency estimation, analyzing its scaling in the thermodynamic limit of small--but finite--Kerr nonlinearity. We provide the parameter set maximizing the QFI, and show that homodyne detection virtually saturates the optimal precision bound. Importantly, the whole analysis considers the role of dissipation in these driven transitions. This allows us to design a highly-sensitive magnetometer, that can be built with state-of-the-art circuit QED technology. 
(ii) Compute the optimal and homodyne-based error probabilities in distinguishing the normal and the symmetry-broken phases. We apply this result to the dispersive qubit readout task in circuit-QED. Our approach goes beyond the semi-classical approximation~\cite{Yamamoto_Kerr,krantz2016single}, and allows one to recognize the set of parameters minimizing the average error probability. We find that the optimal working point lies in proximity of the critical point, in a region where semi-classical or Gaussian approximation can not be applied.

\section*{Results and Discussion}
\subsection*{Kerr resonator model} 
\begin{figure}[h!]
\includegraphics[width=0.9\textwidth]{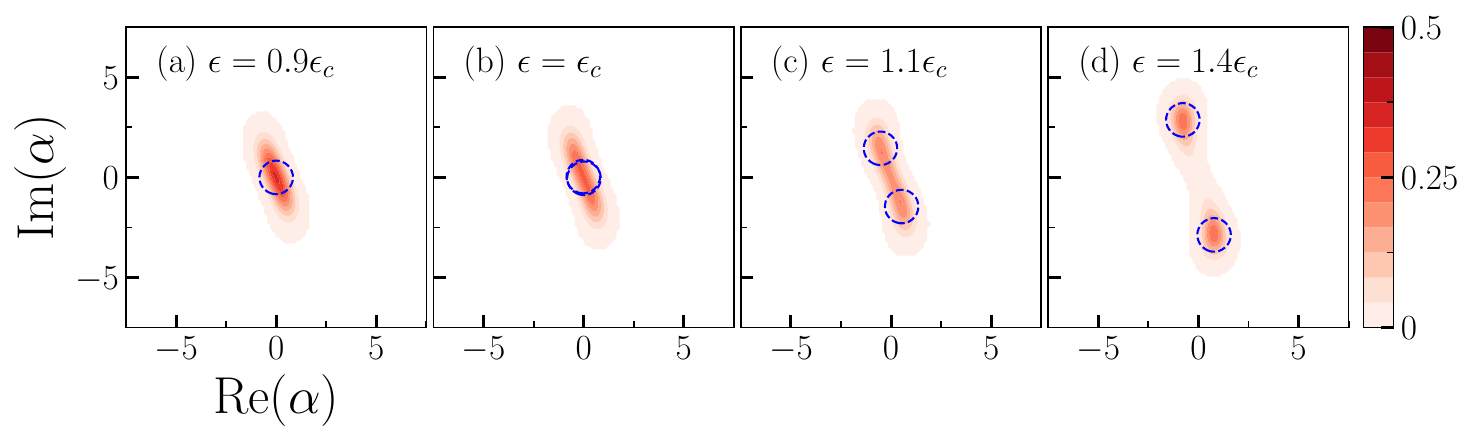} \hspace{1cm}
\caption{\label{Wigner_fig} Wigner function  of the system steady-state obtained with numerical simulations of the full quantum model (colormap), and  half-height contours (dashed blue circles) of the corresponding analytical solutions obtained under semi-classical approximation (see Methods). The four sub plots are obtained taking $\omega = 1~\Gamma$, $\chi = 0.04~\Gamma$ and for increasing values of the pump strength $\epsilon$. The figure shows the transition from the normal (a) to the symmetry-broken [(c) and (d)] phases, taking place around the semiclassical prediction $\epsilon=\epsilon_c = \sqrt{\omega^2+\Gamma^2}$ (b). The system is highly susceptible in the proximity of the criticality, and so it can be exploited in high-sensitivity magnetometry. Moreover, the system shows two highly distinguishable phases, corresponding to a vacuum-like (a) and displaced state (d), a feature that can be exploited in high-fidelity qubit readout.}
\end{figure}

Our starting point is the Kerr-resonator model, whose Hamiltonian is
\begin{equation}
\label{KerrRes}
\hat{H}_{\rm Kerr}/\hbar = \omega \crea{a} \ann{a} + \frac{\epsilon}{2} ( \hat a^\dagger{}^2 +  \ann{a}^2 ) + \chi \crea{a}{}^2\ann{a}^2.
\end{equation} 
This $\mathbb{Z}_2$-symmetric model can be realized in various photonic platforms. In particular, we consider the case a circuit-QED implementation,  where a resonator at frequency $\omega_r$ is coupled with a superconducting quantum interference device (SQUID) element~\cite{Krantz_2013, Yamamoto_Kerr}. If the resonator is pumped at a frequency $\omega_p\simeq 2\omega_r$, then Eq.~\eqref{KerrRes} describes effectively the system, by interpreting  $\omega=\omega_r-\omega_p/2$ as the pump-resonator detuning, $\epsilon$ as the effective pump-power, and $\chi$ as the SQUID-induced nonlinearity. We consider the system embedded in a Markovian thermal bath at zero temperature, described by the Lindblad dissipation superoperator $\mathcal{L}_D[\cdot] = \hbar\Gamma[2\hat a\cdot \hat a^\dag -\{\hat a^\dag\hat a ,\cdot\}]$,
where $\Gamma\geq0$ is loss rate induced by the system-bath coupling. Such a dissipator leaves the model $\mathbb{Z}_2$ invariant~\footnote{See the Supplemental Material.}.
With no loss of generality, we take $\epsilon$ positive. For $\chi=0$, the model is Gaussian and its phenomenology can be easily explained. In the absence of noise, for $\Gamma=0$, the model has a ground state only for $\epsilon<|\omega|$. This is a squeezed vacuum state with squeezing approaching infinity in the $\epsilon/|\omega|\rightarrow 1$ limit. When the bath is turned on, for $\Gamma>0$, the diverging point is shifted. In this case, the steady-state is a squeezed thermal-state and exists only for $\epsilon<\sqrt{\omega^2+\Gamma^2}\equiv \epsilon_c$, with purity approaching zero when $\epsilon/\epsilon_c\rightarrow 1$. The effect of 
the nonlinearity $\chi>0$ is to regularize the model for all parameter values, thus erasing the divergences. In the scaling limit $\chi \rightarrow 0$ a second-order dissipative phase transition (DPT) emerges, associated with the spontaneous breaking of the $\mathbb{Z}_2$-symmetry of the model~\cite{Bartolo2016Exact,Note1}. The steady-state is still Gaussian for $\epsilon<\epsilon_c$. Beyond the critical point, for $\epsilon>\epsilon_c$, the steady-state  is double-degenerate, and it is given by a statistical mixture of two equiprobable displaced squeezed thermal-states~\cite{Minganti2018}, see Fig.~\ref{Wigner_fig}. Since $\chi$ can be made small in a circuit QED implementation, we can exploit the presence of this DPT for both quantum parameter estimation and discrimination. 
On the one hand, we can use the large susceptibility of the steady state in the proximity of the critical point, in order to get a good estimation of $\omega$. In turn, as the resonator frequency has a steep dependence on the external magnetic field threading the SQUID loop, the DPT can be applied in the design of a critical magnetometer. On the other hand, the presence of the DPT allows one to faithfully discriminate between two discrete values of $\omega$, each corresponding to a different phase, in a single-shot measurement. 

\subsection*{Quantum parameter estimation} 
Given an observable $\hat O$, we can define the signal-to-noise ratio (SNR) for estimating the parameter $\omega$ as
\begin{align}~\label{SNR}
S_{\omega}[\hat O]=\frac{|\partial_\omega\langle\hat O\rangle_\omega|^2}{\Delta \hat O_\omega^2},
\end{align}
where $\Delta \hat O_\omega^2=\langle\hat O^2\rangle_\omega-\langle\hat O\rangle_\omega^2$, and the expectation values are computed in the steady-state manifold. This standard definition of SNR we use is useful for parameter estimation protocols because it is directly related to the mean-square error of the estimator\cite{Toth_2014}. The corresponding precision over $M$ measurements is $\Delta \omega^2\simeq [MS_{\omega}]^{-1}$. 
In this paper, we consider the SNR for three important measurements: homodyne, heterodyne and the quantum-mechanical optimal given by the QFI. Homodyne detection consists in projecting on the rotated quadrature operator $\hat x_\varphi=\cos(\varphi)\hat x+\sin(\varphi)\hat p$. Due to the $\mathbb{Z}_2$-symmetry of the system, we consider the observable $\hat x_\varphi^2$, and define the homodyne SNR as $S_{\omega}^{\rm Hom}=\max_{\varphi}S_\omega[\hat x_\varphi^2]$. Heterodyne detection 
corresponds to a noisy measurement of the conjugate quadratures, with outcomes $X$ and $P$. We consider the SNR for the outcome $X^2+P^2$, which can be written as $S_{\omega}^{\rm Het}=|\partial_\omega\langle\hat a\hat a^\dag\rangle_\omega|^2/[\langle\hat a^2\hat a^\dag{}^2\rangle_\omega-\langle\hat a\hat a^\dag\rangle_\omega^2]$, see Methods. Finally, if we maximize the SNR in Eq.~\eqref{SNR} among all the observables, we obtain the QFI: $I_\omega=\max_{\hat O}S_\omega[\hat O]$. This can be expressed as~\cite{Calsamiglia_Chernoff}
\begin{align}\label{fidel}
    I_\omega = \lim_{d\omega \rightarrow0}\frac{8}{d\omega^2}\left[1-\sqrt{F(\rho_\omega,\rho_{\omega-d\omega})}\right],
\end{align}
where $F(\rho_\omega,\rho_{\omega'})=[\text{Tr}\,(\sqrt{\rho_\omega\sqrt{\rho_{\omega'}}\rho_\omega})]^2$ is the fidelity between the steady-states $\rho_\omega$ and $\rho_{\omega'}$.

\begin{figure}[h!]
\includegraphics[width=0.7\textwidth]{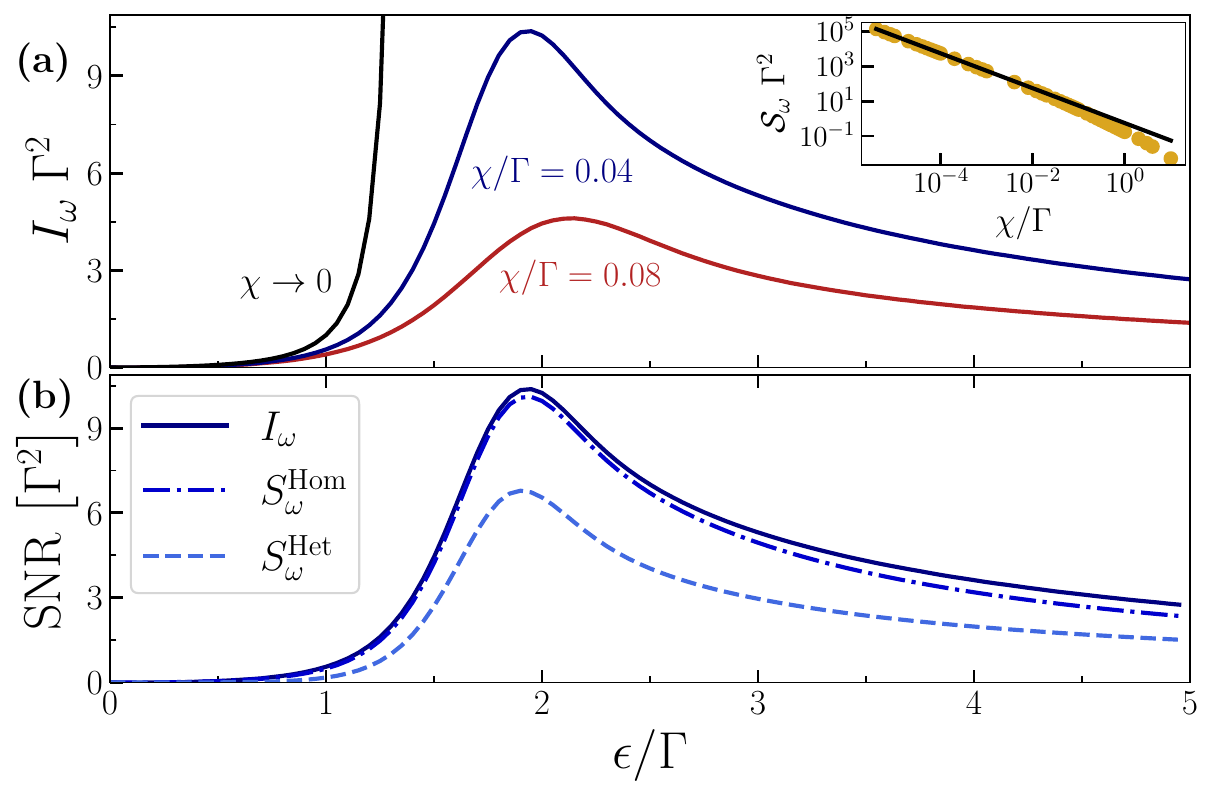} \hspace{1cm}
\caption{\label{SNRfig} {\bf (a)} QFI for the estimation of $\omega$ as a function of $\epsilon$, computed for $\omega/\Gamma=1$ and various values of $\chi/\Gamma$. In the Gaussian case ($\chi\rightarrow0$), the QFI diverges at $\epsilon=\sqrt{\omega^2+\Gamma^2}$. For finite values of $\chi$, the QFI has a maximum. In the inset, we show that $\mathcal{S}_{\omega}=\max_\epsilon S_{\omega}^{\rm Hom}\sim c (\Gamma\chi)^{-1}$, with $c\simeq 0.55$. 
Since $N= \Theta(\sqrt{\chi^{-1}})$, the Heisenberg scaling is reached already for $\chi/\Gamma\lesssim10^{-2}$. {\bf (b)} SNR for the homodyne ($S_\omega^{\rm Hom}$) and heterodyne detection ($S_\omega^{\rm Het}$) at $\omega/\Gamma=1$ and $\chi/\Gamma=0.04$. Homodyne detection virtually saturates the QFI.}
\end{figure}

\emph{The normal phase ($\chi\to0$) -- } 
To begin with, we consider the case $\chi \to 0$, which provides us with a good approximation of the steady-state when we are far enough from the DPT. The model in Eq.~\eqref{KerrRes} with $\chi=0$ has a steady-state solution only for $\epsilon<\epsilon_c$, corresponding to the normal phase.
Using the analytical formula for Gaussian states~\cite{serafini2017quantum}, we compute the QFI with respect to the parameter $\omega$, in the steady-states manifold:
\begin{align}\label{FisherGauss}
I_\omega (\epsilon<\epsilon_c) \xrightarrow[]{\chi\to0} \frac{1}{2\epsilon_c^2-\epsilon^2}\left[2N+\frac{8\omega^2}{\epsilon^2}N^2\right], \end{align}
where $N=\epsilon^2/[2(\epsilon_c^2-\epsilon^2)]$ is the number of photons [see Fig.~\hyperref[SNRfig]{2(a)}].  We have two possible diverging scaling for $\epsilon/\epsilon_c\rightarrow 1$. For $\omega\not=0$ we retrieve the Heisenberg scaling $I_\omega=O(N^2)$,  while for $\omega=0$ one has $I_\omega=O(N)$. Notice that here we focused on the scaling with respect to the number of photons, which is the most relevant figure for the relevant regime of parameters. However, even if the Gaussian model presents a critical slowing down, the Heisenberg scaling can in principle be achieved also with respect to time~\cite{garbe2020,Fully_connected}. We notice also that the divergence rate $I_\omega/N^2$ is maximal at $\omega= \Gamma$. In the following, we focus at this point, where the QFI is maximal for low-enough $\chi$. 

\emph{The symmetry-broken phase ($\chi\to0$) -- } 
The model is invariant under the transformation $\hat{a}\to - \hat{a}$, resulting 
in a $\mathbb{Z}_2$-symmetry.
In the $\chi\to0$ limit, and for $\epsilon>\epsilon_c$, such a symmetry is broken resulting in a second-order DPT. The symmetry-broken solutions are well-approximated by Gaussian states that can be obtained by displacing the field $\hat a\rightarrow \hat a+\alpha$, with $\alpha\in\mathbb{C}$~\cite{felicetti2020universal}.
For nonzero $\chi$, the steady state is well-approximated by a statistical mixture of two Gaussian states~\cite{MingantiSciRep16}. Indeed, a Gaussian approximation leads to $\rho=\frac{1}{2}[D(\alpha) \rho_+ D(\alpha)+D(-\alpha)\rho_- D(-\alpha)]$. 
Here,  $\rho_{\pm}$ are the steady-states for $H_{\pm}= \omega^\prime\hat a^\dagger \hat a + \frac{1}{2}\left(\epsilon^\prime \hat a^\dagger{}^2 + {\epsilon^\prime}{}^* \hat a^2 \right) + O(\sqrt{\chi})$ and dissipator $\mathcal{L}_{D}$,
where $\omega' = 2 \sqrt{ \epsilon^2 -\Gamma^2 } - \omega$ and $|\epsilon'|=\epsilon_c$. 
Namely, $\alpha$ is the solution of  $\omega \alpha + \epsilon \alpha^* +2 \chi |\alpha|^2 \alpha  -i\Gamma\alpha =0 $, see Methods. By setting $\alpha= |\alpha| e^{i\phi}$, we find the two solutions, holding for $\epsilon>\epsilon_c$:
\begin{align}
|\alpha|^2 &= \frac{\sqrt{ \epsilon^2 -\Gamma^2 }  - \omega}{2\chi} \label{alpha1}
, \quad \phi=\frac{\arcsin{\left({\Gamma}/{\epsilon}\right)} \pm \pi }{2}.
\end{align}
Notice that the Hamiltonians $H_\pm$ are the same at the zeroth order in $\chi$. Therefore, $\rho_{+}\simeq \rho_{-}$ and the steady-state solutions consist in a mixture of two identical squeezed-thermal states displaced in opposite directions~\cite{MingantiSciRep16}. 
The QFI shows a divergence at $\epsilon\rightarrow \epsilon_c$, as seen in the normal phase. This confirms that in the proximity of the transition the QFI diverges for $\chi\rightarrow 0$. 
Instead, for sufficiently large $\epsilon$, the QFI value is solely determined by the response of $\alpha$ to the $\omega$'s changes. Using Eq.~\eqref{alpha1}, one can easily see that $I_\omega= \Theta(\epsilon^{-1})$ for $\epsilon\gg1$.

\emph{The full model (finite $\chi$) -- }
We are now ready to show our results beyond the Gaussian approximation. Hereafter, the observables for the QFI are obtained through the analytical solutions in Refs.~\cite{Bartolo2016,MingantiSciRep16, RobertsPRX20}, while the steady-state density matrix are obtained solving the equation $-i [\hat{H}_{\rm Kerr}, {\rho}_{\rm ss}]+\mathcal{L}_{D}[{\rho}_{\rm ss}] =0$ via sparse LU decomposition~\cite{qutip2}. We then compute the QFI using Eq.~\eqref{fidel}. The effect of the Kerr term is to regularize the model, eliminating the divergences that appear in the Gaussian approximation. As expected, the QFI increases with $\epsilon$ up to a maximum point, then it starts to decrease. The maximum  is reached for $\epsilon=\epsilon_c$ in the  $\chi\rightarrow 0$ limit. 
Let us consider the quantity $\mathcal{S}_\omega=\max_{\epsilon}S_\omega^{\rm Hom}$, and focus on the $\omega=\Gamma$ point. With a numerical fit, we find that $\mathcal{S}_{\omega}\simeq c(\chi\Gamma)^{-1}$ for $\chi/\Gamma\lesssim 10^{-2}$, where $c\simeq0.55$ (see Fig.~\hyperref[SNRfig]{2(a)}). Since $N= \Theta(\sqrt{\chi^{-1}})$ holds, the Heisenberg scaling is reached already for $\chi/\Gamma\lesssim10^{-2}$. 
In Fig.~\hyperref[SNRfig]{2(b)}, we show that homodyne detection virtually saturates the QFI already for $\chi/\Gamma=0.04$. In fact, one can easily see that homodyne detection is optimal in the $\chi\to0$ limit, see Methods.

\subsection*{Magnetometry} We now consider an application of our results for the quantum estimation of magnetic flux. Let us consider a SQUID coupled with a $\lambda/4$ resonator. A magnetometer can be designed by coupling the magnetic field into the SQUID loop. The effective Hamiltonian is given in Eq.~\eqref{KerrRes}. Here, the resonator frequency $\omega_r$ depends on the external magnetic flux as $\omega_r(\Phi)\simeq \omega_{\lambda/4}/[1+\gamma_0/|\cos(\Phi)|]$, where $\omega_{\lambda/4}$ is the bare resonant frequency in the absence of the SQUID, $\Phi=\pi \Phi_{\rm ext}/\Phi_0$ is the applied magnetic flux $\Phi_{\rm ext}$ in unit of the flux quantum $\Phi_0$, and $\gamma_0$ is the ratio of SQUID inductance at zero external magnetic flux and the geometric inductance of the resonator.  For $\pi/4 \lesssim \Phi <\pi/2$, where the pump-induced non-linearity is small, the non-linearity depends on the magnetic field as  $\chi(\Phi)\simeq \chi_0\omega_{\lambda/4}\gamma_0^3/|\cos^3(\Phi)|$, where $\chi_0=
\pi Z_0e^2/(2\hbar)$ is a constant dependent on the resonator characteristic impedance $Z_0$~\cite{Note1}. Assuming a typical value $Z_0\simeq50~\Omega$~\cite{Krantz_2013}, we have $\chi_0\simeq 0.02$. It is convenient to work at the point $\Phi\simeq \pi/4$, where $\chi$ is minimized.
We can also assume $\chi$\ to be independent on $\Phi$, by working in the limit $\chi_0\gamma_0^2\ll1$, which ensures the condition $\left|\frac{\partial \chi}{\partial \Phi}\right| \ll\left|\frac{\partial \omega}{\partial \Phi} \right|$ to hold. 
The protocol consists in: (i) Apply a constant magnetic flux bias $\Phi\simeq\pi/4$ to the SQUID. (ii) Apply a pump at frequency $\omega_p\simeq 2[\omega_r(\pi/4)-\Gamma]$. This allows to work at $\omega\simeq\Gamma$, where the QFI is maximal. (iii) Perform homodyne detection of the output signal. 

From the input-output theory, we have that the resonator output mode is $\hat a_{\rm out}=\sqrt{2\Gamma}~\hat a- \hat a_{\rm in}$, where $\hat a_{\rm in}$ is the input mode assumed to be in the vacuum~\cite{gardiner2004quantum}. By applying the right temporal filter at the output mode, one can retrieve the same statistics of the intracavity mode~\cite{Eichler_detection, Ingrid_WignerNegativityKerr}. With this premise, the SNR for the output mode is the same as the one derived for the intracavity mode. 
A change of $\Phi$ by $\delta \Phi$ induces the shift  $\omega\rightarrow \omega+\frac{\partial \omega_r}{\partial \Phi} \delta \Phi$. Therefore, the uncertainty over $M$ independent measurements is $\Delta \Phi_{|\Phi\simeq \pi/4}\simeq \left[\sqrt{
\mathcal{S}_{\omega} M}\left|\frac{\partial \omega_r}{\partial \Phi}_{|\Phi\simeq \pi/4}\right|\right]^{-1}$. Let us consider the regime $\chi/\Gamma\lesssim 10^{-2}$, where $\mathcal{S}_{\omega}\simeq c(\Gamma \chi)^{-1}$, see Fig.~\hyperref[SNRfig]{2(a)}. Let us assume an independent measurement every $2\pi/\Gamma$, and a measurement time of half a second, i.e.  $M=\Gamma$~Hz$^{-1}/(4\pi)$. The magnetometer sensitivity becomes
\begin{align}\label{uncert}
    \frac{\Delta \Phi}{\sqrt{{\rm Hz}}}\lesssim 0.8\left(\frac{\gamma_0}{\omega_{\lambda/4}}\right)^{1/2}
\end{align}
for $\gamma_0\lesssim10^{-2}$, see Methods.
Best sensitivity values reported in the literature are of the order of $4.5\times10^{-7}$~$\sqrt{\text{Hz}^{-1}}$~\cite{sensitivity2017}. Our protocol improves this value by one order of magnitude if we set $\omega_{\lambda/4}\simeq 2\pi\times10$~GHz, and $\gamma_0\simeq 10^{-4}$. The sensitivity can be greatly enhanced by engineering more sophisticated circuit schemes using SQUID arrays or other Josephson-junction configurations, high-impedance metamaterials and large circuits lengths.

\subsection*{Dispersive qubit readout} We now discuss an application of the Kerr resonator for  superconducting-qubit readout.
By dipersively coupling a qubit to the resonator Hamiltonian in Eq.~\eqref{KerrRes}, the Hamiltonian becomes
\begin{align}\label{disp}
    \hat{H}_{\rm disp}/\hbar= \hat{H}_{\rm Kerr}/\hbar+(\omega_r+\Delta)|e\rangle\langle e|+\delta\omega |e\rangle\langle e| \crea{a} \ann{a}.
\end{align}
Here, $\delta\omega=g^2/\Delta$ is a frequency-shift that depends on the qubit-resonator coupling $g$ and the qubit-to-resonator detuning $\Delta$~\cite{Boissonneault_dispersive}. When the qubit is in its excited state $\ket{e}$, a frequency-shift is induced onto the resonator. The Hamiltonian $\hat{H}_{\rm disp}$ can be derived by applying perturbation theory to the full qubit-resonator Hamiltonian, for $g/\Delta\ll1$. The dispersive approximation holds as long as $g^2N/(4\Delta^2)\equiv \eta\ll1$, where $N$ is the number of photons in the resonator. Notice that a small $\eta$ also minimizes the disturbance induced to the qubit by the readout scheme~\cite{Note1}. In the following, we show how the presence of a DPT leads to two highly distinguishable quantum states, that can be used to perform high-fidelity qubit readout.  
A similar setup, with an unoptimized set of parameters, has been experimentally investigated in Ref.~\cite{krantz2016single}. Here, the authors map the qubit discrimination problem to distinguish between the vacuum and a classical state of $\sim 200$~photons, where the dispersive approximation is clearly not valid anymore. As a result, the qubit significantly suffers from additional dissipation processes mediated by the readout resonator. Therefore, one must reduce the number of photons, bringing the system closer to the critical point, where the semiclassical approximation does not hold and  quantum fluctuations shall unavoidably be taken into account. In the following, we conduct this performance analysis in a systematic way using the full quantum model, identifying the set of parameters that maximizes the readout fidelity, while still respecting the dispersive approximation.

\begin{figure}[h!]
\includegraphics[width=0.6\textwidth]{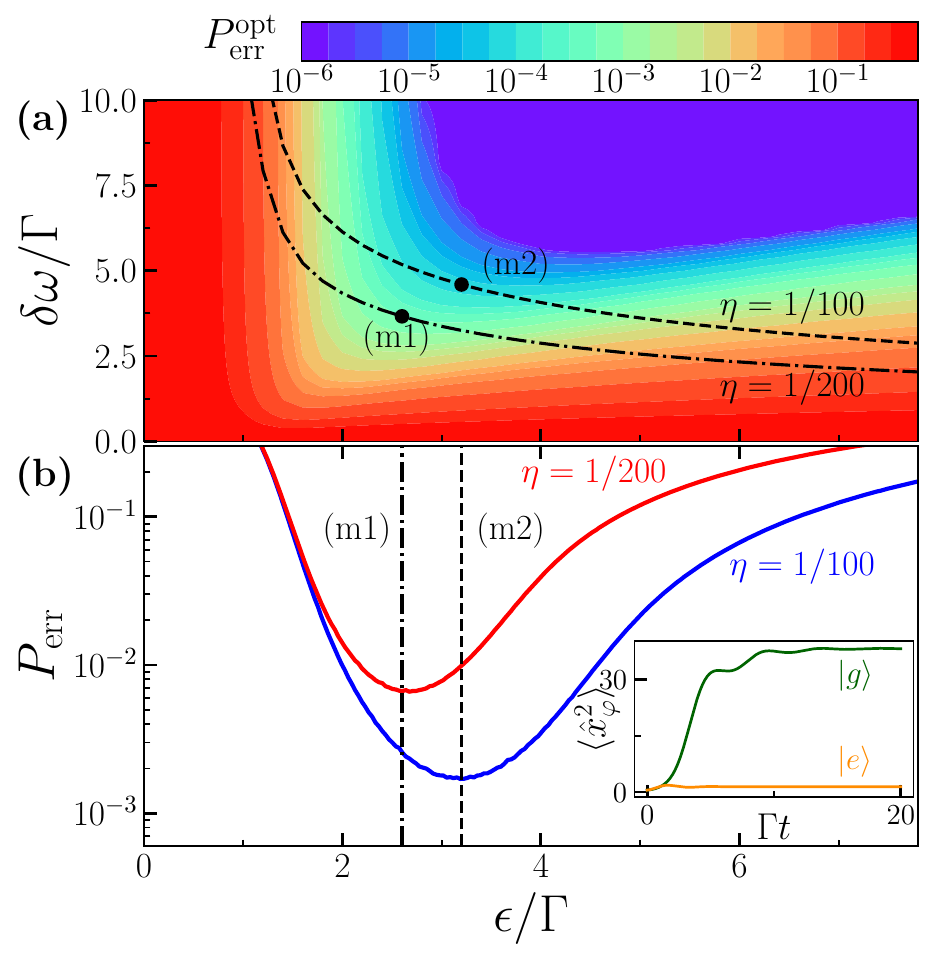} \hspace{1cm}
\caption{\label{Q_readout} {\bf(a)} Error probability map with respect to $\delta \omega/\Gamma=g^2/(\Gamma\Delta)$ and $\epsilon/\Gamma$, for $\omega=0$ and $\chi/\Gamma=0.08$. The dashed lines represent different values of the dispersive parameter $\eta=N\delta \omega^2/(4g^2)$, where $N=\max\{N_{|g\rangle},N_{|e\rangle}\}$ and we have fixed $g/\Gamma=10^2$ to be in the strong---but not ultrastrong---coupling regime. For $\eta=10^{-2}$, we can reach error probability values as low as $10^{-4}$ with the optimal measurement. {\bf (b)} Error probability for homodyne detection at the optimal points and optimal angle $\varphi$ for different values of $\eta$. The inset shows the separation in time of $\langle \hat x^2_{\varphi}\rangle$ for the normal and symmetry-broken phases. The steady-state value is reached at $\Gamma t\simeq10$.} 
\end{figure}

Generally speaking, the method consists in discriminating between two density matrices, i.e. $\rho_g$ and $\rho_e$, corresponding to the steady-states when the qubit is in the state $|e\rangle$ or $|g\rangle$ respectively. We limit ourselves to the case of discrimination via a single measurement of the mode $\hat a$ (single-shot readout). The average error probability is bounded by $P_{{\rm err}}\leq P_{\rm {err}}^{{\rm opt}}$, also known as the Helstrom bound~\cite{Calsamiglia_Chernoff}, where  $P_{\rm {err}}^{{\rm opt}}= \frac{1}{2}\left[1-\frac{1}{2}\|\rho_e-\rho_g\|_1\right]$
and $\|\sigma\|_1=\text{Tr}\,\sqrt{\sigma^\dag\sigma}$ is the trace norm~\footnote{The qubit readout fidelity can be defined as $F=1-P_{\rm err}$.}. The optimal error probability $P_{\rm {err}}^{{\rm opt}}$ is in principle achievable by measuring in the eigenbasis of $\rho_e-\rho_g$. In Fig.~\hyperref[Q_readout]{3(a)} we show a map of the $P_{\rm {err}}^{{\rm opt}}$ values with respect to the frequency-shift $\delta \omega$ and the pump strength $\epsilon$, for $\chi=0.08\Gamma$. For a given value of $\eta$, the graph shows the presence of a sweet spot where the error probability is minimized. The value $P_{\rm {err}}^{{\rm opt}}$ is always attainable, and gives us a bound on what error probabilities can be in principle reached. However, the measurement can be complicated to implement, so we consider also a practical strategy  based on  homodyne detection. Let us define the probability density functions $P_{g,e}(x)=\int W_{g,e}(x,p)dp$, where $W_{g,e}(x,p)$ are the Wigner functions of the resonator steady-state in the case of qubit in the $|g\rangle$ or $|e\rangle$ states. We declare that the state of the qubit is $|g\rangle$ if our measurement outcome belongs to $\{x|P_g(x)>P_e(x)\}$ and $|e\rangle$ otherwise. The error probability for this discrimination strategy is 
\begin{align}
    P_{\rm err}=\frac{1}{2}\int\min_x \{P_g(x),P_e(x)\}dx.
\end{align}
This procedure can be further optimized by considering a rotated homodyne measurement. Notice that we have used the Wigner function as a tool to find the best threshold value distinguishing between the two qubit states, given a single homodyne measurement. Indeed, our strategy does not rely on the reconstruction of the resonator Wigner function. In Fig.~\hyperref[Q_readout]{3(b)}, we show that the homodyne strategy, although not saturating the optimal strategy, achieves error probability values of the order of $10^{-3}$. Notice, however, that the position of the minimum of the error probability with the homodyne strategy coincides with that of $P_{\rm {err}}^{{\rm opt}}$. 
In the Supplemental Material, we show that for $\eta=10^{-2}$ there is no backaction on the qubit states~\cite{Note1}.
Experimentally achievable values attaining the optimal value for  $\eta=10^{-2}$ are:  $\Gamma\simeq2\pi\times 1$~MHz, $\chi/\Gamma\simeq0.08$, $g/\Gamma\simeq 10^2$, $\omega_r/\Gamma\simeq 8\times 10^3$, $\omega_q/\Gamma\simeq 6\times 10^3$. In this case, the resonator has at most $N\simeq 30$ photons at the steady-state.

\section*{Methods}

\subsection*{The Gaussian approximation for $\chi \rightarrow 0$}\label{suppl:I}
\vspace{0.3cm}

Here, we find the Gaussian approximation for $\chi\to0$, for both the regimes $\epsilon<\epsilon_c$ (normal phase) and $\epsilon>\epsilon_c$ (symmetry-breaking phase).
\vspace{0.5cm}

\emph{The normal phase ($\epsilon<\epsilon_c$) -- }
For $\epsilon<\epsilon_c$, we set $\chi=0$ and look for the steady-state solutions. It is convenient to rewrite the master equation as Fokker-Planck equation, in the Wigner function formalism:
\begin{equation}
\label{FPlack}
\frac{\partial W}{\partial t}(x,p)=-(\omega - \epsilon)p\frac{\partial W}{\partial x} + (\omega +\epsilon) x\frac{\partial W}{\partial p}+\Gamma\left[ 2W+\sum_{i=1}^2 x_i\partial_iW +
\frac{1}{2} \sum_{i=1}^2\partial^2_i W \right], 
\end{equation}
where we defined $\hat a = (\hat x+i\hat p)/\sqrt{2}$. 
Since this equation is quadratic in $x$ and $p$, it can be solved by a Gaussian ansatz  
\begin{equation}
W=\frac{1}{\pi \sqrt{\det \sigma}} \exp\Curly{-\frac{1}{4} \sum_{i,j} r_i(\sigma^{-1})_{ij}r_j},
\end{equation}
where $r = ( x, p )$ and the Wigner function is normalized to one. The covariance matrix $\sigma$ is defined as $\sigma_{ij} = \langle \{\hat r_i , \hat r_j\} \rangle - 2 \langle \hat r_i \rangle \langle \hat r_j \rangle$. 
From \eqref{FPlack} we get the linear system of equations
\begin{align}
\label{equationsigmabis}
\partial_t\sigma & = B\sigma+\sigma B^T- 2\Gamma(\sigma - \sigma^L) \, 
\end{align}
where $ \sigma^L = \mathbb{I}_2$ and $
B =\begin{bmatrix}0 & (\omega - \epsilon) \\ - (\omega + \epsilon) & 0\end{bmatrix}$. 
We find the steady state by solving $\partial_t \sigma_{\rm ss} = 0$. The solution is
\begin{align}\label{sigmasteady}
\sigma_{\rm ss}=\frac{1}{\epsilon_c^2-\epsilon^2}
\begin{bmatrix}
\epsilon_c^2 - \omega \epsilon &   - \Gamma \epsilon \\
- \Gamma \epsilon & \epsilon_c^2  + \omega\epsilon
\end{bmatrix},
\end{align}
which corresponds to a physical state only for $\epsilon^2 < \omega^2 + \Gamma^2\equiv\epsilon_c^2$. This sets the critical value in the $\chi \rightarrow 0$ limit. In this limit, the number of photons is 
\begin{align}
N(\chi\to0)=\frac{(\sigma_{\rm ss})_{11}+(\sigma_{\rm ss})_{22}-2}{4}=\frac{\epsilon^2}{2(\epsilon_c^2-\epsilon^2)},
\end{align}
which diverges for $\epsilon\to \epsilon_c$.
\vspace{0.5cm}

\emph{The symmetry-broken phase ($\epsilon>\epsilon_c$) -- }
Let us derive an effective quadratic Hamiltonian for $\epsilon>\epsilon_c$. We follow the approach developed in \cite{felicetti2020universal}. The idea is that for small $\chi$ the model is well approximated by a double-well potential, and that the low-energy physics can be described with a quadratic expansion around each minimum.  In order to center the reference frame on one of the two minima, let us apply a displacement operation such that $U^\dagger \hat a U = \hat a + \alpha $.  We obtain an effective Hamiltonian
\begin{equation}\label{displ}
\hat{H}_\alpha = \hat{H}^{(1)} + \hat{H}^{(2)} + \hat{H}^{(3/4)} + \text{const.}, 
\end{equation}
where
\begin{eqnarray}\label{H1}
\hat{H}^{(1)}&=&\left( \omega \alpha + \epsilon \alpha^* +2 \chi |\alpha|^2 \alpha -i\Gamma \right) \hat a^\dagger + \text{H.c.} \\
\label{H2}
\hat{H}^{(2)} &=& \left( \omega + 4 \chi |\alpha|^2 \right) \hat a^\dagger \hat a  +  \left( \frac{\epsilon}{2} + \chi \alpha^2 \right) {\hat a^\dagger}{}^2 +  \left( \frac{\epsilon}{2} + \chi {\alpha^*}^2 \right) \hat a^2 , \\ \label{H3}
\hat{H}^{(3/4)} &=& \chi\left( \hat a^\dagger \hat a^\dagger \hat a \hat a + 2 \alpha {\hat a^\dagger}{}^2 \hat a + 2 {\alpha^*} \hat a^\dagger \hat a^2 \right).
\end{eqnarray} 
The dissipator $\mathcal{L}_D$, instead, is left unchanged. The quadratic part of the displaced Hamiltonian  \eqref{displ} is well-defined, i.e. it has normal modes with positive frequency and is bounded from below. Accordingly, far from the critical point the steady state will have bounded quantum fluctuations, and the norm of the creation/annihilation operators on the steady state will be bounded. In the limit of small $\chi$, and of large $\alpha$, higher-order terms are negligible and the model is well-approximated by a Gaussian approximation which includes only terms quadratic in $\crea{a}$ and $\ann{a}$ [see the solutions below in Eq. \eqref{sol_alpha}]. Of course this approximation will break in a small region for $\epsilon \rightarrow \epsilon_c^+$, and the size of the critical region is proportional to $\chi$ (the smaller the nonlinearity, the more reliable the Gaussian approximation even as the critical point is approached).

  The linear equation defining the equilibrium points is found by imposing $\hat{H}^{(1)}=0$, i.e. $\omega \alpha + \epsilon \alpha^* +2 \chi |\alpha|^2 \alpha - i \Gamma \alpha = 0$. Setting $\alpha= |\alpha| e^{i\phi}$ we find two solutions for $\epsilon>\epsilon_c$:
\begin{eqnarray}
\label{sol_alpha}
|\alpha|^2 &=& \frac{\sqrt{ \epsilon^2 -\Gamma^2 }  - \omega}{2\chi} \label{alphasuppl} \\ 
\phi &=& \frac{1}{2}\arcsin{\left(\Gamma/\epsilon\right)} \pm \pi/2.\label{phisuppl}
\end{eqnarray}
We find the effective Hamiltonians in the symmetry-broken phase by plugging the solution into Eq.~\eqref{displ}. We get
\begin{equation}
H_\pm = \omega^\prime a^\dagger a + \frac{1}{2}\left(\epsilon^\prime {a^\dagger}^2 + {\epsilon^\prime}^* a^2 \right) + O(\sqrt{\chi}), 
\end{equation}
with
\begin{align}
   \omega^\prime &= 2 \sqrt{ \epsilon^2 -\Gamma^2 } - \omega \\
   \epsilon^\prime &=\epsilon_c e^{i\theta} \\
   \theta &=- 2\arctan\left[\frac{\Gamma\left(\sqrt{\epsilon^2-\Gamma^2}-\omega\right)}{\epsilon\epsilon_c+\omega\sqrt{\epsilon^2-\Gamma^2}+\Gamma^2}\right].
\end{align}
Notice that the Hamiltonians $H_\pm$ are the same for $\chi \rightarrow 0$. Therefore, in this limit the two solutions are degenerate. Increasing the pump power $\epsilon$ corresponds to an effective growth of the pump-resonator detuning, since $\omega^\prime\sim \epsilon$ for large $\epsilon$. Instead, the effective squeezing parameter $\epsilon'$ remains constant in modulus, while its argument changes until reaching the value $\theta_{\epsilon/\epsilon_c\gg1}=-2\arctan\sqrt{(\epsilon_c-\omega)/(\epsilon_c+\omega)}$. Therefore, the effect of increasing the pump is to displace the state to the new equilibrium points, and to rotate and reduce the squeezing of each of the resulting states.
\vspace{1cm}

\subsection*{Quantum parameter estimation}
\vspace{0.3cm}

Here, we derive the quantum parameter estimation results for the full model. 
\vspace{0.5cm}

\emph{Signal-to-noise ratio (SNR) -- }
The SNR induced by the observable $\hat O$ in the task of estimating the parameter $\omega$ is defined as
\begin{equation}\label{SNRdef}
S_\omega[\hat O]=\frac{[\partial_\omega \langle \hat O\rangle_\omega]^2}{\Delta \hat O^2_\omega},
\end{equation}
where $\Delta \hat O^2_\omega=\langle \hat O^2\rangle_\omega-\langle \hat O\rangle_\omega^2$ and the index $\omega$ indicates the expectation value computed on the steady-state $\rho_\omega$. The SNR computed in $\omega=\omega_0$ should be interpreted as the precision achievable for estimating the parameter $\omega$ when its value is close to $\omega_0$, through the relation $\Delta \omega^2_{|\omega\simeq \omega_0}\simeq [M \times S_{\omega_0}]^{-1}$, where $M\gg1$ is the number of measurements. Generally speaking, if an experimentalist is able measure the expectation value of a class of observables $\{O(\vec r)\}$,  with ${\vec r}=(r_1,\dots,r_K)$, they would like to maximize the SNR with respect to ${\vec r}$ in order to obtain a better precision rate (call $\vec r_{\rm max}$ the maximizing set of parameters). This in principle requires the preknowledge of $\omega_0$. If this knowledge is not provided, then they can implement a two-step adaptive protocol, where first they measure the expectation value of $A\in\{O(\vec r)\}$ such that the function $f(\omega)=\langle A\rangle_\omega$ is invertible in the range of values where $\omega$ belongs, obtaining a first order estimation of $\omega$, i.e. $\omega_0$. Then they find $\vec r_{\rm max}$ and measure $O(\vec r_{\rm max})$.

We are particularly interested in the following SNRs.
\begin{itemize}
    \item {\it Homodyne detection}: This is defined by the POVM $\mathcal{X}_\varphi^{\rm Hom} = \{\ket{x_\varphi}\bra{x_\varphi}\}_{x_\varphi\in\mathbb{R}}$, where $|x_\varphi\rangle$ is an eigenstate of the rotated quadrature 
    \begin{align}
    \hat x (\varphi) = \cos(\varphi)\hat x + \sin(\varphi)\hat p  = \frac{1}{\sqrt{2}}[\hat a e^{-i\varphi} + \hat a^\dagger e^{i\varphi}].
     \end{align}
    We consider this SNR $S_\omega[\hat x_\varphi^2]$ to evaluate the perfomance of homodyne detection for finite $\chi$. This will be clear in the following, when we will evaluate the classical Fisher information. We also consider the best phase choice, i.e.
    \begin{align}\label{SNRHom}
        S_\omega^{\rm Hom}\equiv \max_\varphi S_\omega[\hat x_\varphi^2].
    \end{align}

    \item {\it Heterodyne detection}: This is defined by the POVM $\mathcal{X}^{\rm Het}=\left\{\frac{1}{\pi}|\gamma\rangle\langle \gamma|\right\}_{\gamma\in\mathbb{C}}$, where $|\gamma\rangle$ is a coherent state. The heterodyne measurement can be modeled as the signal $\hat a$ entering in a beamsplitter with a thermal mode $\hat h$ as the other input, obtaining the two modes $\hat b_{\pm}=[\hat a\pm\hat h^\dag ]/\sqrt{2}$. The quadratures $\hat X_+=[\hat b_++\hat b_+^\dag]/\sqrt{2}$ and $\hat P_-=-i[\hat b_--\hat b_-^\dag]/\sqrt{2}$ are finally measured. This is equivalent to measure the complex envelope operator 
\begin{equation}\label{complexenv}
\hat S\equiv \hat X_++i\hat P_-=\hat a+\hat h^\dag.
\end{equation}
Indeed, from the measurement outcomes one can estimate the moments of $\hat S$, and then invert Eq.~\eqref{complexenv} to obtain an estimation of the moments of $\hat a$. In the quantum-limited case, when  $\hat h$ is a vacuum mode, the moments of $\hat S$ can be easily computed  as $\langle \hat S^{\dag m}\hat S^n\rangle=\langle \hat a^n \hat a^{\dag m}\rangle$, since $[\hat S,\hat S^\dag]=0$~\cite{daSilvaCircuitQEDdetection, Di_Candia_2014DualPath}. The SNR for the observable $\hat O_{\rm Het}=\hat X_+^2+\hat P_-^2=\hat S^\dag \hat S$ is then
\begin{align}\label{SNRHet}
S_\omega[\hat O_{\rm Het}]=\frac{[\partial_\omega \langle \hat a\hat a^\dag\rangle_\omega]^2}{\langle \hat a^2\hat a^{\dag}{}^2\rangle_\omega-\langle\hat  a \hat a^\dag\rangle_\omega^2}\equiv S_\omega^{\rm Het}.
\end{align}

\item {\it Classical Fisher information (FI)}: Let us focus on the homodyne POVM 
$\mathcal{X}^{\rm Hom}_\varphi = \{\ket{x_\varphi}\bra{x_\varphi}\}_{x_\varphi\in\mathbb{R}}$. The estimation precision is given by the FI, which is defined as the SNR maximized over the observables which are diagonal in $\{\ket{x_\varphi}\bra{x_\varphi}\}_{x_{\varphi}\in\mathbb{R}}$, i.e. $F_\omega(\mathcal{X}_\varphi^{\rm Hom})=\max_{\hat O=\sum_{x_\varphi}p(x_\varphi)|x_\varphi\rangle\langle x_\varphi|}S_\omega[\hat O]$. This can be generally expressed as~\cite{Paris_quantum_estimation}
\begin{equation}
F_\omega(\mathcal{X}_\varphi^{\rm Hom}) = \int_\mathbb{R} dx\ p(x_\varphi|\omega)\ \left\{ \partial_\omega \left[ \ln\ p(x_\varphi|\omega) \right] \right\}^2,
\end{equation}
where $p(x_\varphi|\omega) = \bra{x_\varphi} \rho_\omega \ket{x_\varphi}$ is the probability density function of the outcome.

\item {\it Quantum Fisher information (QFI)}: Generally speaking, the QFI provides the precision for the optimal unbiased estimator allowed by quantum mechanics. This is indeed defined as $I_\omega=\max_{\hat O} S_\omega[\hat O]$, that can be generally expressed as~\cite{Calsamiglia_Chernoff}
\begin{align}\label{QFIdef}
    I_\omega = \lim_{d\omega \rightarrow0}\frac{8}{d\omega^2}\left[1-\sqrt{F(\rho_\omega,\rho_{\omega-d\omega})}\right],
\end{align}
where $F(\rho_\omega,\rho_{\omega-d\omega})=\left(\text{Tr}\,\sqrt{\rho_\omega\sqrt{\rho_{\omega-d\omega}}\rho_\omega}\right)^2$ is the fidelity between the two density matrices $\rho_{\omega}$ and $\rho_{\omega-d\omega}$, and $\rho_\omega$ ($\rho_{\omega-d\omega}$) is the steady-state of the model with pump-resonator detuning $\omega$ ($\omega-d\omega$). This expression can be easily evaluated numerically by considering a $d\omega$ smaller and smaller, and by doing a convergence check. In the paper, this procedure has been used to compute the QFI in the steady-state manifold for the $\chi>0$ case. Indeed, Eq.~\eqref{QFIdef} has been evaluated numerically by letting the system evolve to the steady-state for two values of $
\omega$ close to each other. 
\end{itemize}
\vspace{0.5cm}

\subsection*{Quantum parameter estimation for $\chi\rightarrow 0$ (Normal phase)}
\vspace{0.3cm}

Here, we compute the QFI and the FI for the Gaussian model. We show that homodyne detection saturates the QFI for $\chi\to0$.

\emph{QFI for the Gaussian model --}
The QFI for the Gaussian model can be analytically calculated using the covariance matrix formalism~\cite{serafini2017quantum}, using the solutions of Eq.~\eqref{sigmasteady}:
\begin{equation}
I_\omega = \frac{1}{2 (1+\mu^2)} \text{Tr} \left[ \left( \sigma^{-1} \partial_\omega \sigma \right)^2 \right]\ +\ \frac{2 (\partial_\omega \mu)^2 }{1-\mu^4},
\end{equation}
where $\mu =(\det \sigma)^{-1/2}$ is the purity of the quantum state, and we have used that the displacements are zero for all parameter values. The solution has been computed using a symbolic computation software, obtaining
\begin{align}
I_\omega = \frac{\epsilon^2}{2\epsilon_c^2-\epsilon^2}\left[\frac{1}{\epsilon_c^2-\epsilon^2}+\frac{2\omega^2}{(\epsilon_c^2-\epsilon^2)^2}\right].
\end{align}
This expression can be cast as in Eq.~\eqref{FisherGauss} using the relation $N=\epsilon^2/[2(\epsilon_c^2-\epsilon^2)]$.
We recall that the quantum Fisher information provides an upper bound to the achievable SNR as defined in Eq.~\eqref{SNR}, which can be saturated when the optimal measurement is implemented.

\emph{FI for the Gaussian model -- }
Let us focus on the homodyne  POVM 
$\mathcal{X}^{\rm Hom}_{\varphi} = \{\ket{x_\varphi}\bra{x_\varphi}\}_{x_\varphi\in\mathbb{R}}$. 
One can easily see that in the Gaussian approximation  $ F_{\omega}(\mathcal{X}_\varphi^{\rm Hom})= \left[ \partial_\omega s_\omega(\varphi) \right]^2/(2 s_\omega(\varphi)^2)= S_\omega[\hat x_\varphi^2]$, where $s_\omega (\varphi) = \cos^2(\varphi)\ (\sigma_{ss})_{11} + \sin^2(\varphi)\ (\sigma_{ss})_{22} - \sin(2\varphi)\ (\sigma_{ss})_{12}$~\cite{bina_dicke_2016}. This can be expanded as
\begin{align}
    F_\omega(\mathcal{X}_\varphi^{\rm Hom})= \frac{\epsilon^2\left[(\Gamma^2-\omega^2-\epsilon^2)\cos(2\varphi)+2\omega\epsilon+2\omega\Gamma\sin(2\varphi)\right]^2}{2(\epsilon_c^2-\epsilon^2)^2[\epsilon_c^2-\epsilon(\omega\cos(2\varphi)-\Gamma\sin(2\varphi))]^2}.
\end{align}

\begin{itemize}
\item For $\omega=0$, we have that 
\begin{align}
    F_{\omega=0}(\mathcal{X}_\varphi^{\rm Hom})=\frac{\epsilon^2\cos^2(2\varphi)}{2[\epsilon_c^2+\epsilon_c\epsilon\sin(2\varphi)]^2}.
\end{align}
This is maximal for $\sin(2 \varphi)=-\epsilon/\epsilon_c$, for which we obtain $F_{\omega=0}(\mathcal{X}_\varphi^{\rm Hom})=\epsilon^2/[2\epsilon_c^2(\epsilon_c^2-\epsilon^2)]$. Since $F_{\omega=0}(\mathcal{X}_\varphi^{\rm Hom})/I_{\omega=0}\rightarrow 1/2$ for $\epsilon/\epsilon_c\rightarrow1$, we have that homodyne does not saturate the QFI when $\omega=0$. However the FI and the QFI share the same diverging scaling, i.e. $F_{\omega=0}(\mathcal{X}_\varphi^{\rm Hom})=O(\epsilon_c-\epsilon)^{-1}$ for $\varepsilon\to \varepsilon_c$. 

\item For $\omega\not=0$, the FI scales differently with respect to $\epsilon_c-\epsilon$. We exemplify the calculation for $\omega=\Gamma$, that is the point where the QFI shows the maximal divergence rate. Here, we have that 
\begin{align}
    F_{\omega=\Gamma}(\mathcal{X}_\varphi^{\rm Hom})=\frac{\epsilon^2[\epsilon_c^2\sin(2\varphi)-\epsilon^2\cos(2\varphi)+\sqrt{2}\epsilon_c\epsilon]^2}{\epsilon_c^2(\epsilon_c^2-\epsilon^2)^2[\sqrt{2}\epsilon_c-\epsilon(\cos(2\varphi)-\sin(2\varphi))]^2}.
\end{align}
In the $\epsilon/\epsilon_c\rightarrow1$ limit, we have that $ F_{\omega=\Gamma}(\mathcal{X}_\varphi^{\rm Hom})\sim \epsilon_c^2/(\epsilon_c^2-\epsilon^2)^2$ for any $\varphi$. This means that $F_{\omega=\Gamma}(\mathcal{X}_\varphi^{\rm Hom})/I_{\omega=\Gamma}\rightarrow 1$, and homodyne detection saturates the QFI for any $\varphi$,
\end{itemize}
\vspace{0.5cm}

\subsection*{Quantum parameter estimation for $\chi>0$}
\vspace{0.3cm}

Let us evaluate the QFI scaling in two regimes: (i) $\epsilon/\epsilon_c \gg1$, and (ii) $\epsilon$ close to the criticality. 

\begin{itemize}
\item  $\epsilon/\epsilon_c\gg1$. In this regime, we have seen that the Wigner function becomes a mixture of two equiprobable coherent states, symmetrically displaced with respect to the center. These states are uniquely determined by $|\alpha(\omega)|^2$, as the phase $\phi$ in Eq.~\eqref{phisuppl} does not depend on $\omega$. Therefore, for symmetry reasons, the optimal observable is the photon-number operator. This gives rise to the optimal SNR scaling
\begin{align}
    I_\omega \sim S_\omega[\hat N]&=\frac{\left|\partial_\omega (|\alpha|^2)\right|^2}{|\alpha|^2}=\frac{1}{2\chi\left[\sqrt{\epsilon^2-\Gamma^2}-\omega\right]}\sim \frac{1}{2\chi\epsilon}.
\end{align}

\item  {\it $\epsilon$ close to the criticality.} In this case, we have analyzed numerically the scaling of $\mathcal{S}_\omega=\max_\epsilon S_\omega^{\rm Hom}$, computed in $\omega=\Gamma$, where the QFI is maximal for low-enough $\chi$. We find that $\mathcal{S}_{\omega}(\omega=\Gamma)\sim c(\chi\Gamma)^{-1}$ for $\chi/\Gamma\lesssim0.01$, where $c\simeq0.55$. In addition, since in the same regime we have that $N= \Theta(\sqrt{\chi^{-1}})$, that the Heisenberg scaling is reached. Let us focus on $\mathcal{I}_\omega=\max_\epsilon I_\omega$. On the one hand, we always have that $\mathcal{I}_{\omega}(\omega=\Gamma)\geq \mathcal{S}_{\omega}(\omega=\Gamma)$. On the other hand, in practice homodyne detection already saturates the QFI for $\chi/\Gamma=0.04$, meaning that one should expect $\mathcal{I}_{\omega}(\omega=\Gamma)\simeq \mathcal{S}_{\omega}(\omega=\Gamma)$ already in this regime, since homodyne performs optimally for $\chi\to0$.
\end{itemize}

\vspace{1cm}

\subsection*{The magnetometer sensitivity}
\vspace{0.3cm}

Here, we derive the sensitivity of the magnetometer in Eq.~\eqref{uncert}. We have that 
\begin{align}
    \Delta \Phi_{|\Phi\simeq\pi/4}\simeq\left[\sqrt{\mathcal{S}_{\omega}(\omega\simeq\Gamma) M}\left|\frac{\partial \omega_r}{\partial \Phi}_{|\Phi\simeq \pi/4}\right| \right]^{-1},
\end{align}
where $\mathcal{S}_{\omega}(\omega=\Gamma)\simeq c(\chi\Gamma)^{-1}$ for $\chi/\Gamma\lesssim0.01$, $M=\Gamma~{\rm Hz}^{-1}/(4\pi)$, and $\omega_r(\Phi)\simeq \omega_{\lambda/4}/[1+\gamma_0/|\cos(\Phi)|]$. In the $0\leq\Phi\leq \pi/2$ regime, we have that 
\begin{align}
    \frac{\partial \omega_r}{\partial \Phi} \simeq -\frac{\gamma_0\omega_{\lambda/4}\sin(\Phi)}{(\gamma_0+\cos(\Phi))^2}.
\end{align}
It follows that 
\begin{align}
 \frac{\Delta \Phi_{|\Phi\simeq\pi/4}}{\sqrt{\rm Hz}} &\simeq \sqrt{\frac{\pi}{2c}}\frac{(2\gamma_0+\sqrt{2})^2}{\gamma_0}\frac{\sqrt{\chi(\Phi\simeq\pi/4)}}{\omega_{\lambda/4}}.
\end{align}
We now use that $\chi(\Phi)\simeq\chi_0\omega_{\lambda/4}\gamma_0^3/|\cos^3(\Phi)|$ to obtain
\begin{align}
    \frac{\Delta \Phi_{|\Phi=\pi/4}}{\sqrt{\rm Hz}} &\simeq 2^{1/4}\sqrt{\frac{\chi_0}{4\pi^3c}}(2\gamma_0+\sqrt{2})^2\sqrt{\frac{\gamma_0}{\omega_{\lambda/4}}}\\
    \quad &\simeq0.39(2\gamma_0+\sqrt{2})^2\sqrt{\frac{\gamma_0}{\omega_{\lambda/4}}},
    \end{align}
where we have used that $\chi_0=
2\pi^3Z_0e^2/\hbar\simeq 0.02$ for $Z_0\simeq50~\Omega$, and $c\simeq0.55$. This means that
\begin{align}
   \frac{\Delta \Phi_{|\Phi=\pi/4}}{\sqrt{\rm Hz}}\lesssim 0.8 \sqrt{\frac{\gamma_0}{\omega_{\lambda/4}}},
\end{align}
for $\gamma_0\lesssim10^{-2}$, which is Eq.~\eqref{uncert}.

Finally, let us see how $\chi$ changes with respect to small changes of $\Phi$. In the $0\leq\Phi\leq \pi/2$ regime, we have that 
\begin{align}
 \frac{\partial \chi}{\partial \Phi}\simeq 3\chi_0\omega_{\lambda/4}\gamma_0^3\frac{\sin(\Phi)}{\cos^4(\Phi)}.
\end{align}
At $\Phi=\pi/4$, the condition $\left|\frac{\partial \chi}{\partial \Phi}\right| \ll\left|\frac{\partial \omega}{\partial \Phi} \right|$ is equivalent to $6\chi_0\gamma_0^2\ll1$.
\vspace{1cm}

\subsection*{Dispersive qubit readout}
\vspace{0.3cm}

\emph{Dispersive Hamiltonian -- }
Consider a qubit-resonator system, with Hamiltonian 
\begin{align}
    \hat{H}_{\rm qr}=\hat{H}_{\rm JC}+ \frac{\hbar\epsilon}{2}(\hat a^2+\hat a^\dag{}^2)+\hbar\chi\hat a^\dag{}^2\hat a^2, 
\end{align}
where $\hat{H}_{\rm JC}/\hbar=\omega_r\hat a^\dag \hat a+\omega_q|e\rangle\langle e|+ g(\sigma^-\hat a^\dag +\sigma^+\hat a)$ is the Jaynes-Cumming Hamiltonian with qubit-resonator coupling $g$  and qubit-frequency $\omega_q$. Dispersive-readout protocols assume the qubit-resonator coupling to be in the linear dispersive regime, where an effective Hamiltonian is found applying the unitary transformation $U=\exp\{(g/2\Delta)[\sigma^+\hat a-\sigma ^-\hat a^\dag]\}$, where $\Delta=|\omega_q-\omega_r|$ is the qubit-resonator detuning,  and applying perturbation theory with respect to $g/\Delta$. One then finds the effective Hamiltonian
\begin{align}\label{disp1}
    \hat{H}_{\rm disp}/\hbar= \omega_q|e\rangle\langle e|+(\omega+\delta\omega |e\rangle\langle e|) \crea{a} \ann{a} + \frac{\epsilon}{2} ( \hat a^{\dagger^2} +  \ann{a}^2 ) + \chi \crea{a}{}^2\ann{a}^2,
\end{align}
that can be cast as in Eq.~\eqref{disp}. Here, $\delta\omega=g^2/\Delta$ is a qubit-state dependent frequency-shift, that depends on the qubit-resonator coupling $g$ and the qubit-resonator detuning $\Delta$~\cite{Boissonneault_dispersive}. The dispersive approximation holds as long as $g^2N/(4\Delta^2)\equiv \eta\ll1$, where $N$ is the number of photons in the resonator.
\vspace{0.5cm}

\emph{Optimal parameter choice -- }
For each sample, the state of the resonator collapse either on $\rho_e$ or $\rho_g$. Discriminating between these two states gives us the measurement result. Fixing a value of $\chi/\Gamma$, one can draw a ($\delta\omega,\epsilon$)-dependent map of the optimal error probability for discriminating $\rho_e$ and $\rho_g$, i.e. $P_{\rm err}^{\rm opt}=\left[1-\|\rho_e-\rho_g\|/2\right]/2$. Since $\epsilon$ is monotone with respect to $N$, for each value of $(\delta\omega, \epsilon, \eta)$, one can find a value of $g$ and $\Delta$ satisfying the conditions $\delta \omega=g^2/\Delta$ and $\eta=g^2N/(4\Delta^2)$. However, since $g/\Gamma$ cannot be too large, otherwise we go to the ultrastrong regime where the counter-rotating terms appear, we have fixed $g/\Gamma=10^2$, and choose $\omega_r$ and $\omega_q$  such that  $g/\min\{\omega_q,\omega_r\}\ll1$. We have then drawn the lines for $\eta$ equals to $10^{-2}$ and $0.5\times 10^{-2}$, see Fig.~\hyperref[Q_readout]{3(a)}.

\bibliographystyle{apsrev4-1}
\bibliography{References.bib}

\section*{Acknowledgements}
R.D. acknowledges support from the
Marie Sk{\l}odowska Curie fellowship number 891517 (MSC-IF Green-MIQUEC) and the Academy of Finland grants no. 353832 and 349199. K.P. and G.S.P. acknowledge the funding from the European Union’s Horizon 2020 European Union’s Horizon 2020 Research and Innovation Action under grant agreement No. 862644 (FET-Open project: Quantum readout techniques and technologies, QUARTET). We are grateful to the Academy of Finland for support through the RADDESS grant No. 328193 and through the ``Finnish Center of Excellence in Quantum Technology QTF” grants Nos. 312296, 336810, and 352925.

\section*{Author Contributions}
R. D. and S. F. suggested the idea of the paper. R. D. and S. F. derived the analytical results. F. M. derived the numerical results. K.P. and G.S.P. provided experimentally realistic parameters for the applications. R. D., F. M. and S. F. wrote the manuscript. All authors contributed to discussions and proofreading of the manuscript.


\newpage

\section*{SUPPLEMENTAL MATERIAL}
In this Supplemental Material, we provide additional information for the discussion in the main text. Our starting point is the Kerr-resonator model in Eq.~(1) in the main text. We include the interaction with a bosonic bath at temperature $T=0$, described by a Markovian master equation in Lindblad form:
\begin{equation}\label{Eq:Lindblad_Parametric}
\dot{\rho} = -i\hbar\left[\omega \crea{a} \ann{a} + \frac{\epsilon}{2} ( \hat a^{\dagger^2} +  \ann{a}^2 ) + \chi \crea{a}{}^2\ann{a}^2
, \rho\right] + \hbar\Gamma \left (2\hat a \rho \hat a^\dagger - \left\{ \hat a^\dagger \hat a, \rho \right\}\right),
\end{equation}
where $\Gamma\geq0$ is the resonator dissipation rate, $\chi\geq0$ is the non-linear interaction coupling, and we assume $\epsilon\geq0$ without loss of generality.

\section{Circuit model}
First, we provide more details on the physical origin of the system parameters, in the case of an implementation with a superconducting parametric oscillator. In particular we consider the device sketched here in Supplementary Figure~\ref{sketch}, that is, a $\lambda/4$ resonator grounded through a SQUID.
A thorough experimental characterization of this device has been provided in~\cite{Krantz_2013}, and similar designs have been used in various practical applications such as, for example, in the dispersive readout~\cite{krantz2016single} of superconducting qubits.
 
\begin{figure}[h]
    \centering
    \includegraphics[height=0.30\textwidth]{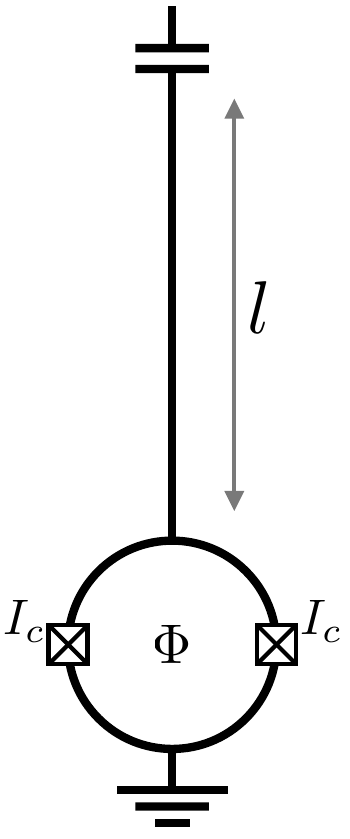}
    \caption{ \label{sketch} Sketch of the circuit scheme. A $\lambda/4$ resonator is grounded through a SQUID device. A  magnetic flux bias $\Phi$ is used to tune the frequency shift and the nonlinearity induced on the resonator.}
\end{figure} 
 
 The frequency of the bare resonator, that is the frequency the system would have if the waveguide was grounded directly to the ground, is given by $\omega_{\lambda/4} = 1/\sqrt{LC}$. Here $L = L_0 l$ is the total inductance, given by the inductance per unit length $L_0$ of the waveguide multiplied by the resonator length $l$. Similarly, $C= C_0 l$ is the total capacitance. The presence of the SQUID introduces a renormalization of the resonator frequency and different kinds of nonlinearities.  The frequency of the device fundamental mode will depend on the external magnetic field $\Phi_{ext}$ threading the device, and it is well approximated by~\cite{Krantz_2013}
\begin{equation}
\omega_r(\Phi) \simeq \frac{\omega_{\lambda/4}}{1 + \frac{\gamma_0}{\cos(\Phi)}},
\end{equation}
where $\Phi = \pi\frac{\Phi_{ext}}{\Phi_0}$ and $\Phi_0$ is the magnetic flux quantum.
The dependence of the device frequency on the external field depends on the  the parameter $\gamma_0 = \frac{L_s^0}{L} = \frac{\Phi_0}{2\pi I_c} \frac{1}{L}$,  which corresponds to the ratio between the inductance of the SQUID at zero bias and the geometric inductance of the resonator. Here, $I_c$ is the critical current of the Josephson junction forming the SQUID, which are assumed to be symmetrical without loss of generality. 

We will limit ourself here to the Duffing nonlinearity, which is the most relevant in the considered parameter regime. The nonlinear parameter $\chi(\Phi)$ also depends on the external magnetic flux and it can be well approximated by 
\begin{equation}
\chi(\Phi) = \chi_0 \left( \frac{\gamma_0}{\cos(\Phi)} \right)^3, \quad \text{where} \qquad \chi_0 = \frac{\pi \omega_{\lambda/4} Z_0}{R_K},
\end{equation}
where we defined the resonator geometric impedance $Z_0 = \sqrt{L/C}$ and the quantum resistance $R_K = h/e^2$. 

\section{Second-order dissipative phase transition in the two-photon Kerr resonator}
We provide now some details on the phase transition of the two-photon driven Kerr resonator. We focus on the case in which the pump frequency is red-shifted with respect to the resonator, where the transition is of second order. That is, the system reaches the same steady state when the critical point is approached from the normal and superradiant phases, and no hysteretical phenomena take place. For a more complete discussion, we refer the interested reader to Refs.~\cite{Bartolo2016Exact,Savona2017,minganti_spectral_2018,Rota2019}. 
At first, let us remark that Eq.~\eqref{Eq:Lindblad_Parametric} is characterized by a $\mathbb{Z}_2$-symmetry. Indeed, any transformation $\hat{a}\to - \hat{a}$ leaves the equation unchanged. While for closed systems the presence of a symmetry implies a conserved quantity, this is not always the case for open-quantum systems~\cite{Albert2014}. Nevertheless, even if in the dissipative case the  $\mathbb{Z}_2$-parity is not necessarily preserved during the dynamical evolution, the symmetry constraints the properties of the steady-state density matrix. Indeed, one can show that $\operatorname{Tr}[\hat{a} {\rho}_{\rm ss}]= 0 $. 
For finite-size systems (i.e. when no scaling on the system parameters or on the number of particles is performed), one has that ${\rho}_{\rm ss} = \lim_{t \to \infty} {\rho}(t)$. Indeed, the steady state of this model is unique, and any initial state will eventually converge to it.
The transition in the parametric resonator is the emergence of multiple steady states that ``break'' the symmetry, and $\lim_{t \to \infty} \operatorname{Tr}[\hat{a} {\rho}(t)]\neq 0 $.
As such, we can define two phases: the \textit{normal}, and the \textit{symmetry broken} phases. 

\subsection{The thermodynamic limit}
As we just discussed, no phase transition can occur unless an infinite scaling is imposed on the system parameters. For extended systems, such as lattices of size $L$, criticality emerges in the thermodynamic limit of infinite sites ($L\to \infty$). In a single Kerr resonator, one can instead replace the thermodynamic limit with a scaling on the system parameters. Under this rescaling of physical parameters the system enters a regime where the system ground (or steady) state becomes populated by a large number of photons. Accordingly, the system spans an increasingly large region of the infinite-dimensional bosonic Hilbert space, and so it reproduces the thermodynamic limit of infinite particles which is defined for extended or many-body systems. 
Several arguments, such as semiclassical analysis~\cite{Bartolo2016Exact}, G\"utzwiller mean-field approximations~\cite{Savona2017}, or Bogoliubov-like approximations~\cite{minganti_spectral_2018} indicate that the scaling towards the thermodynamic limit can be obtained by introducing a scaling parameter $L$ such that
\begin{equation}
    \langle \hat{a}^\dagger \hat{a} \rangle (L) \to \frac{\langle \hat{a}^\dagger \hat{a} \rangle}{L}, \quad \chi(L) = \frac{\chi_0}{L}.
\end{equation}
Although a true phase transition can be reached only in the thermodynamic limit, sizable effects can be witnessed also for finite-size systems, i.e. for large but finite $L$, as discussed in Ref.~\cite{minganti_spectral_2018}. 
We show the scaling towards the phase transition in Supplementary Figure~\ref{phasetrans}. While ``far'' from the thermodynamic limit (small $L$) the curves are very different one from the other, in the large-$L$ limit they converge to the same value, showing a universal behavior in the thermodynamic limit. 
Changing $\omega$ has a two-sided effect. On the one hand, the transition point is shifted [compare Figs.~\hyperref[phasetrans]{2(a)} and \hyperref[phasetrans]{(b)}], which means that the transition takes place for different values of the pump intensity. On the other hand, $\omega$ affects the shape of the states around the transition. Notice that this dependence on $\omega$ can be seen only by considering the full quantum model. Indeed, no dependence on $\omega$ of the system steady state is observed under the semiclassical approximation (that corresponds to assuming that the oscillator is in a classical coherent state) or using a Gaussian model (which corresponds to a quadratic expansion of the system Hamiltonian).

\subsection{The role of the nonlinearity}
\begin{figure}
    \centering
    \includegraphics[width=0.90\textwidth]{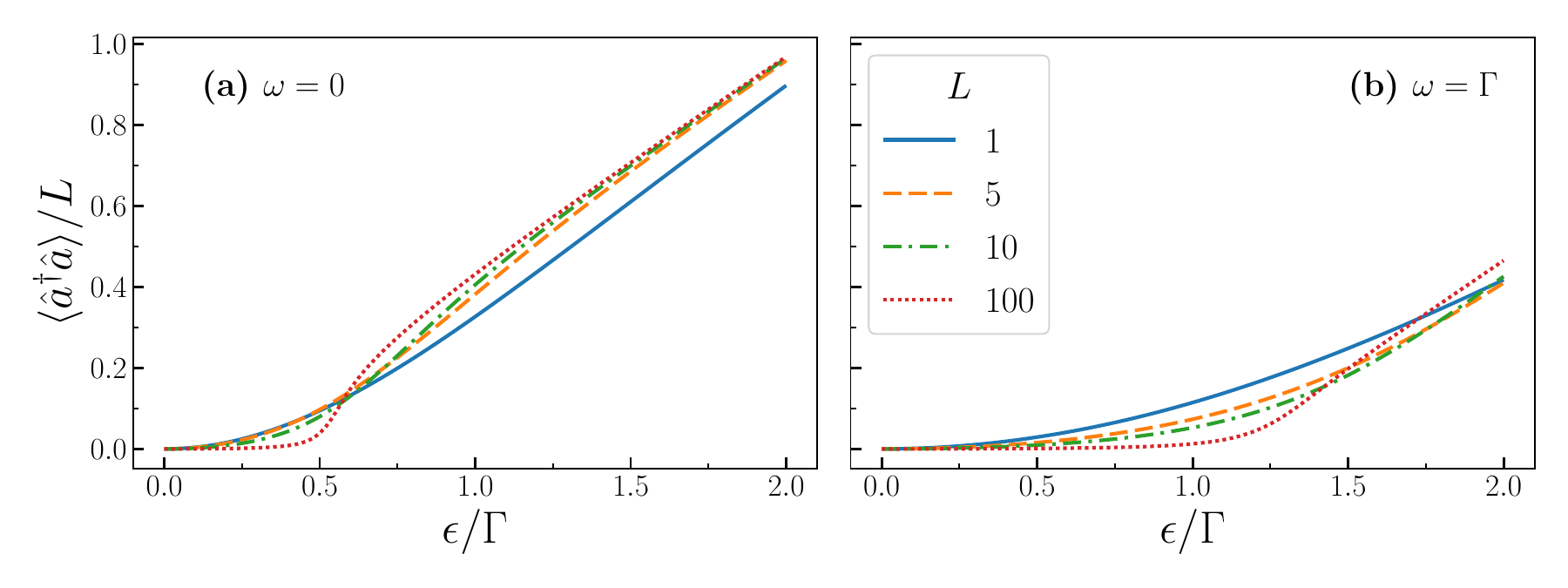}
    \caption{ \label{phasetrans} Onset of the second-order phase transition in the parametric Kerr resonator for  $\omega=0$ [panel (a)] and $\omega=\Gamma$ [panel (b)]. Different curves represent different values of the rescaling parameter $N$, such that $\chi\rightarrow\chi/L$, the thermodynamic limit being reached for $L\to \infty$. The nonlinearity is fixed at $\chi_0/\Gamma=1$.}
\end{figure}

Before proceeding further, let us remark here the fundamental role of the nonlinearity in this model.
Since $\chi/L \to 0$, one would be tempted to remove $\chi$ from the equation of motion in the thermodynamic limit.
While this approach is justified in the normal phase, this is not the case in the symmetry-broken case. Indeed, an optical parametric resonator without a nonlinearity is known to undergo a parametric instability~\cite{carmichael_statistical2}.
For instance, consider Supplementary Figure~\hyperref[phasetrans]{2(a)} in the ``symmetry-broken'' phase for $\epsilon/\Gamma\simeq 2$. The curves for $L=5$, $L=10$, and $L=100$ overlap with each other.
For these curves, the effect of the nonlinearity compared to that of other operators, e.g., the dissipation, can be estimated as
\begin{equation}
    \frac{\chi(L) \langle\hat{a}^\dagger{}^2 \hat{a}^2 \rangle}{\Gamma  \langle \hat{a}^\dagger \hat{a} \rangle} \simeq  \frac{\chi_0/L}{\Gamma} \langle \hat{a}^\dagger \hat{a} \rangle \simeq \frac{\chi_0}{\Gamma} \frac{\langle \hat{a}^\dagger \hat{a} \rangle}{L} \simeq \frac{\chi_0}{\Gamma} \langle \hat{a}^\dagger \hat{a} \rangle (L).
\end{equation}
In other words, even if we increase the value of $L$, the effect of $\chi$ is constant. 
As such, even a very small nonlinearity can produce a sizeable effect on the characteristics of the transition.

\section{Readout backaction and dispersive approximation validity}

There can be several backaction effects on the qubit due to the measurement apparatus: 
\begin{enumerate}
	\item The cavity-qubit coupling changes the qubit parameters, inducing, for instance, a shift in the qubit frequency.
	\item The cavity-qubit coupling is exactly described by the dispersive approximation only in the limit of vanishingly small light-matter interaction or for infinitely-detuned fields. As such, very small effects beyond the dispersive approximation can slightly modify the qubit and cavity dynamics with respect to the simplified description used to  modelize the qubit-readout protocol.
	\item The fact that the overall readout process must behave as a measurement inevitably implies that
	the qubit wave function must collapse in one of the eigenstates of $\hat{\sigma}_z$.
\end{enumerate}

Concerning (a), the shift in the qubit parameters due to the presence of the cavity can easily be  taken into account and eliminated by appropriately tuning the cavity-to-pump detuning.

As for (b), in the regime we considered, effect beyond those described by this Hamiltonian are of the order of $\eta^2$, and therefore negligible for the level of analysis presented here.
For this reason, the dispersive coupling considered in the text is a valid approximation. 
We have also verified numerically that, for the ``optimal'' point considered in the text, the results of the simulation with full qubit-to-cavity coupling changes less that $0.5\%$ when compared with the approximated one.

Finally, concerning point (c), one can easily show that our protocol is indeed inducing a projective measurement on the qubit, despite the fact that $\eta \ll1$.
To demonstrate this fact, we consider that the system is initialized in
\begin{equation}
\ket{\Psi(t=0)} = \frac{\ket{e}+\ket{g}}{\sqrt{2}} \otimes \ket{0},
\end{equation}
that is, the qubit is pointing in the positive direction along $\hat{\sigma}_x$, and the cavity is empty.
Accordingly, we expect that $1/2$ of the time the outcome is $\ket{e}$, the other half  $\ket{g}$, so that in average, $\langle \hat{\sigma}_z \rangle =0$. To simulate this process, we resort to single homodyne quantum trajectories~\cite{carmichael_statistical2},  i.e., we simulate a single realization of an ideal experiment in which the cavity output field is probed via a perfect homodyne detection.
We show the results of this simulation in Supplementary Figure~\ref{traj}.
\begin{figure}[h!]
\centering
	\includegraphics[width=0.45 \textwidth]{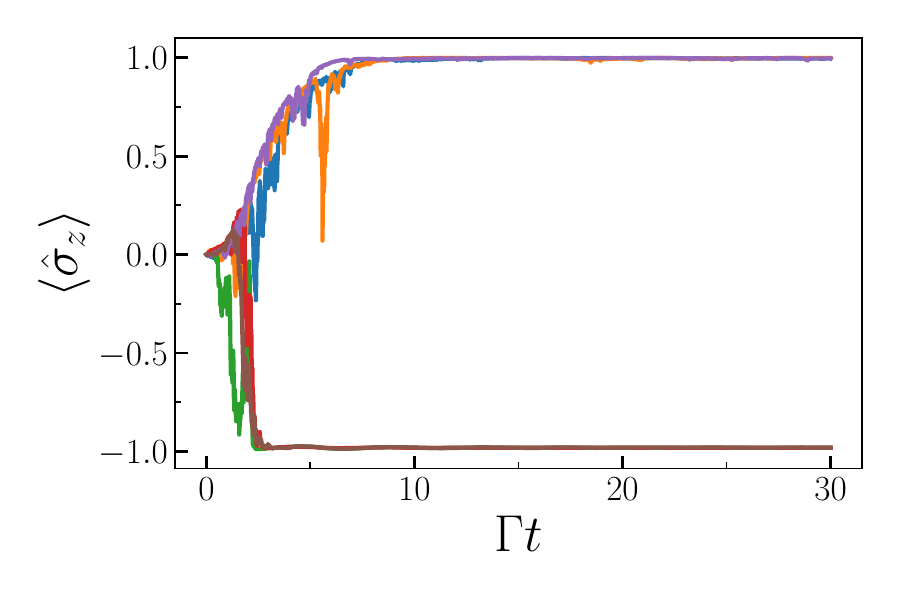}
	\includegraphics[width=0.45 \textwidth]{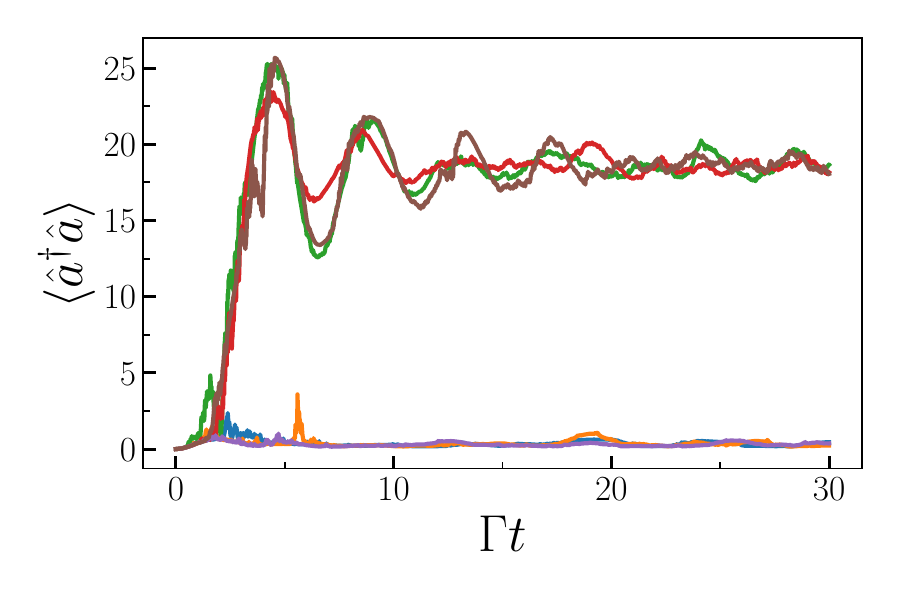}
	\caption{Single homodyne trajectories, representing the time evolution of the qubit coupled to a cavity, assuming $\eta=10^{-2}$. Different colors represent different trajectories. As one can see, the overall effect of the measurement is to project the qubit either in the state $\ket{e}$ or in $\ket{g}$ in a short time, comparable to the photon lifetime. Parameters as in Figure~3 in the main text.}\label{traj}
\end{figure}
As one can clearly see, in a very short time the state collapses either in $\ket{e}$ or in $\ket{g}$. In the case under consideration, out of the 6 trajectories 3 collapsed in $\ket{e}$, and 3 in $\ket{g}$.
In other words, the fact that $\eta$ is very small does not mean that the cavity is not having any back-action on the qubit.  Indeed, the high-rate at which the escaping photons are measured rapidly collapse the qubit wave function.
This proves that $\eta=10^{-2}$ is small enough for the dispersive approximation to hold, and to avoid any sort of backaction on the qubit.

\end{document}